\documentclass[journal]{IEEEtran}
\usepackage[colorlinks,urlcolor=blue,linkcolor=blue,citecolor=blue]{hyperref}
\usepackage{color,array}
\usepackage{cite}
\usepackage{amsmath,amssymb,amsfonts,mathpazo}
\usepackage{graphicx, tabularx}
\usepackage{textcomp}
\usepackage{caption}
\usepackage{subcaption}
\usepackage{wrapfig}
\usepackage{algorithm, algorithmic}
\usepackage{xurl}
\usepackage{tikz}
\usepackage{booktabs,tabulary,lipsum}
\usepackage{dcolumn}
\usepackage{siunitx}
\usepackage{multirow}
\usetikzlibrary{shapes, arrows.meta, positioning, calc, shadows}
\usepackage{makecell}
\usepackage[style=american]{csquotes}
\ifCLASSINFOpdf
\else
\fi
\begin{document}
%
\title{Concurrent Learning of Policy and Unknown Safety Constraints in Reinforcement Learning} 
%
%
%

\author{Lunet~Yifru and
        Ali~Baheri
\thanks{Lunet Yifru is with the Department
of Mechanical, Materials and Aerospace Engineering, West Virginia University, Morgantown, WV 26506, USA. email: \tt\small lay0005@mix.wvu.edu}
\thanks{Ali Baheri is with the Department
of Mechanical Engineering, Rochester Institute of Technology, Rochester,
NY 14623, USA. email: \tt\small akbeme@rit.edu}}

\maketitle

\begin{abstract}
Reinforcement learning (RL) has revolutionized decision-making across a wide range of domains over the past few decades. Yet, deploying RL policies in real-world scenarios presents the crucial challenge of ensuring safety. Traditional safe RL approaches have predominantly focused on incorporating predefined safety constraints into the policy learning process. However, this reliance on predefined safety constraints poses limitations in dynamic and unpredictable real-world settings where such constraints may not be available or sufficiently adaptable. Bridging this gap, we propose a novel approach that concurrently learns a safe RL control policy and identifies the unknown safety constraint parameters of a given environment. Initializing with a parametric signal temporal logic (pSTL) safety specification and a small initial labeled dataset, we frame the problem as a bilevel optimization task, integrating constrained policy optimization, using a Lagrangian-variant of the twin delayed deep deterministic policy gradient (TD3) algorithm, with Bayesian optimization for optimizing parameters for the given pSTL safety specification. Through experimentation in comprehensive case studies, we validate the efficacy of this approach across varying forms of environmental constraints, consistently yielding safe RL policies with high returns. Furthermore, our findings indicate successful learning of STL safety constraint parameters, exhibiting a high degree of conformity with true environmental safety constraints. The performance of our model closely mirrors that of an ideal scenario that possesses complete prior knowledge of safety constraints, demonstrating its proficiency in accurately identifying environmental safety constraints and learning safe policies that adhere to those constraints.
\end{abstract}


%
\IEEEpeerreviewmaketitle

\section{INTRODUCTION}

RL has risen as a key computational paradigm involving training intelligent agents to make sequential decisions, aiming to maximize some notion of expected return \cite{sutton2018reinforcement}. It has been instrumental in solving complex dynamic problems across a wide range of applications such as autonomous driving, robotics, aviation, finance, etc. \cite{kiran2021deep, kober2013reinforcement,razzaghi2022survey,hambly2023recent}. However, deploying RL in practical settings introduces the critical concern of safety, especially in domains, such as autonomous driving and healthcare, where unsafe actions can lead to catastrophic outcomes. Safety refers to the need for systems to operate within acceptable risk parameters, and pertaining to RL, safety is defined by the system's ability to attain the environmental objectives while adhering to safety constraints.  

Traditionally, safe RL methods base policy design on either modifying the optimality criterion to include cost as one of the objectives or altering the exploration process \cite{Garcia2015safeRLreview}. For instance, a prominent approach is the integration of formally defined safety constraints, such as STL, into reward functions encapsulating critical limits within which RL agents must operate. This approach is particularly appealing, because, unlike classical machine learning models, which are often black-box and obscure, temporal logic formalism offers a precise, human-interpretable language for system behavior. Logically constraining RL has shown promise for generating safe, high-performance policies, however, the effectiveness of this approach hinges on the availability and quality of the predefined safety constraints. Defining such safety constraints can be effectively approached through the utilization of expert knowledge, manually designed by domain experts, or derived from data using computational techniques. However, the reliance on expert knowledge for defining safety constraints can be restrictive and often infeasible, as experts are not always readily available. Furthermore, given the dynamic nature of environments, experts may have a limited perspective, potentially leading to safety constraints that do not fully encapsulate the true environmental conditions resulting in either overly conservative behavior or overlooked risks. On the other hand, computational approaches for mining temporal logic safety specifications depend on the availability of extensive historical datasets, which may not always be accessible, or its acquisition could pose significant risks in safety-critical domains. Overall, specifying exact safety constraints in RL environments is a challenging task, and static, predefined constraints may not be sufficiently adaptable to address the complexities of dynamic real-world environments.



Conventional safe RL methods fall short in designing safe policies in the absence of predefined safety constraints, leaving a critical gap that impedes the broader integration of safe RL into areas lacking such predefined constraints. To bridge this gap, we propose an approach that enables the learning of safe control policies in environments where safety constraints are not explicitly defined \textit{a priori}. Our approach, given a parametric STL (pSTL) specification and two categories of small initial datasets, one populated with safe trajectories and another populated with unsafe trajectories, \emph{concurrently} identifies the pSTL safety parameters that accurately model the environmental constraints and derives an optimal safe RL policy constrained by the learned STL. Our concurrent learning process is facilitated through the input of a human expert who iteratively provides labels to rollout traces generated by executing the learned policy. This allows the extension of the small initial dataset and efficient refining of the pSTL parameter values, steering them towards the accurate constraints, in turn, guiding the RL algorithm towards the optimal safe policy. 

\noindent{\textbf{Our Contribution.} Key contributions of our paper are: 
\begin{enumerate}
\renewcommand{\labelenumi}{\arabic{enumi}.}
    \item We propose a novel framework for concurrently learning safe RL policies and STL safety constraint parameters in an environment where safety constraints are not defined \textit{a priori}.
    \item  We modify the TD3-Lagrangian constrained RL algorithm to use STL as a constraint specification during policy synthesis.
    \item Through extensive evaluations in various safety-critical environments, and comparisons to baseline models, we prove that our framework is able to obtain safe RL policies that maximize rewards while upholding safety constraints, performing comparably to baseline models equipped with predefined safety constraints. 
\end{enumerate}

The remainder of this paper is structured as follows. Section \ref{sec:Related Work} provides a review of related work in the domain of safe RL and STL synthesis, Section \ref{sec:Preliminaries} outlines the foundational concepts used in deriving our proposed approach. The problem statement is articulated in Section \ref{sec:problem statement} and our methodology is detailed in section \ref{sec:Methodology}. Section \ref{sec: results and discussion} is dedicated to the performance evaluation of our results as compared with baselines and discussion of the implications of our findings as well us the limitations of our work. 
%
%
%
%

\section{Related Work}\label{sec:Related Work}

Our work is related to two key areas of research, namely safe RL policy synthesis and formal safety specification learning with additional emphasis on parameter synthesis of pSTL specifications.

\noindent {\textbf{Safe RL.}} In recent years, a diverse range of approaches for safe RL has been proposed, including constrained RL \cite{achiam2017constrained}, safety layers or shielding \cite{alshiekh2018safe,baheri2020deep,baheri2022safe}, and formal methods \cite{bansal2022specification}. A comprehensive overview of safe RL methods is given in \cite{safeRLreview}. Primal-dual policy optimization \cite{qiu2020UpperConfidencePrimal-Dual, chow2017riskconstrainedPrimal-Dual}, a method prominent to our approach, is based on the Lagrangian relaxation procedure and solves a saddle-point problem to iteratively optimize the policy (primal) while adjusting the dual variable. Taking inspiration from primal-dual methods, constrained policy optimization methods outlined in \cite{achiam2017constrained, yang2020pcpo} develop trust region methods, which approximately enforce constraints in every policy update by evaluating the constraint based on samples collected from the current policy.

Safe RL methods based on using safety certificates as constraints are also explored. An example of such methods is to constrain the agent’s actions by applying the control law of Lyapunov functions, mathematical approaches guaranteeing stability and safety, and then excluding unsafe actions from the action set\cite{Chow2019LyapunovbasedSP, chow2018lyapunovbased}. Lyapunov method, however, requires prior knowledge of a Lyapunov function, which can be challenging to obtain. Another form of safety certificates are Barrier functions that divide unsafe states and safe states by finding a barrier and starting from a given initial state, ensure that the system will not enter the unsafe set. Some works use barrier certificates as constraints \cite{luo2022learningbarriercertificates, yang2023saferlbarriercertificates}.

Another direction explored in a safe RL framework is utilizing Gaussian Process (GP) models. For instance, the SNO-MDP framework optimizes costs within a safe region and maximizes rewards in areas with undefined safety constraints by using GP models to predict unobserved states \cite{SNO-MDP}. Similarly, some research have used GP models to approximate unknown functions for safe exploration \cite{turchetta2016safeFiniteMDPGP}, and to represent unknown reward and cost functions, ensuring safety with a certain probability and optimizing reward\cite{2018wachisafeGP} . 

Conversely, other approaches synthesize safe policies based on reward shaping techniques informed by temporal logic formulae as constraints. For instance, a safe RL method using linear temporal logic (LTL) as a constraint during policy generation has been suggested in \cite{hasanbeig2020cautious}. Other methods of modifying the reward function involve replacing the reward function in RL environments with the robustness degree of an STL constraint \cite{Hamilton, li2017RLWTLrewards}, partial signal rewarding mechanism based on the robustness of a given safety STL specification \cite{BNHR}, the $\tau$-CMDP approach that uses Lagrangian relaxation to solve a constrained optimization by using an STL specification as a constraint \cite{Ikemoto_2022DRLunderSTLandLagrangian}. Although temporal logic-based methods deliver impressive safety performance, the logical constraints need to be \emph{predetermined} to ensure their success. 




\noindent \textbf{Safety Specification Mining.} The learning of STL specification can be divided into two categories: learning of the formula template along with the parameters, and learning of the parameters given the formula template/pSTL specifications. 

Recently research directions have focused on mining complete STL specifications (both template and parameters) from data. The work in \cite{2014kongTLI}, by defining a partial order over the set of reactive STL (rSTL), proposes a passive learning approach that infers an STL specification which serves as a classifier from positive and negative examples. This approach was later extended to an online setting\cite{kong2017temproal} and an unsupervised approach \cite{jones2014AnomalyCPS}. The authors in \cite{2017Vaidyanathangridbased} propose another passive learning method that uses grid-based signal discretization, clustering of similar signals by similarity of covered cells, translating clusters into equivalent STL formulas, and constructing an STL at the disjunction of cluster STL formulas. Decision trees are another widely explored alternative for STL mining and they could be based on offline supervised learning from positive and negative examples \cite{2016BombaraDecisiontreeOffline, 2021BombaraDecisiontreeOnlineanfdoffline}, online supervised learning \cite{2018BombaraDecisiontreeOnline}, and offline unsupervised learning \cite{2017BombaraDecisiontreeofflineunsupervised}. An approach for mining STL specifications from positive examples guided by robustness metrics is introduced by the work in \cite{telex}. Evolutionary algorithms have also been explored in this domain \cite{ROGE,telex,2021pigozzigrammarbased}, using genetic operators to evolve candidate formulas into effective classifiers.

Several approaches have been explored, specifically targeting the parameter synthesis of pSTL specifications. While the computation of the exact validity domain of a pSTL specification has been investigated to address parameter synthesis \cite{asarin2011exactvalidity}, it is evident that this method incurs exponentially increasing computational costs. To mitigate this, this method is extended towards approximating validity domains using run-time verification methods in tandem with search techniques by the same authors in \cite{asarin2011exactvalidity}. The method outlined in \cite{telex} proposes a passive learning of pSTL parameters from positive examples by introducing the notion of a differentiable tightness metric for STL specification satisfaction, and uses gradient-based methods to search over the parameter space. The STLCG framework bridges pSTL parameter mining with machine learning by presenting a novel integration of computation graphs from the machine learning domain to evaluate the robustness of STL formulas and learn pSTL parameters \cite{STLCG}. Conversely, the works in \cite{Jin2015miningrequirements,yang2012querying} adopt an active learning strategy to mine pSTL parameters using signals generated by dynamic models through an iterative process that computes candidate STL specifications and utilizes falsification methods to search for counterexamples generated by the model. This approach is limited in that it necessitates the availability of a dynamic model capable of generating new signals. The method of logical clustering combines pSTL parameter inference with unsupervised learning \cite{VazquezChanlatte2016LogicalCA}. It projects signals to template parameters within their validity domain, uses clustering to group similar signals, and defines an STL formula for each cluster. For our approach, we took inspiration from the ROGE framework, where the parameter identification is addressed using Bayesian optimization (BO)\cite{ROGE}. 

Recent research directions suggest methods for learning safety certificates, and similarly to our proposed approach, some simultaneously learn safe control policies and safety certificates \cite{ma2022joint, luo2022learningbarriercertificates}. However, to our knowledge, there are no works on concurrently learning a temporal logic based safety specification and a constrained RL control policy.

\section{Preliminaries}\label{sec:Preliminaries}
\subsection{Signal Temporal Logic} \label{ssec:STLPreliminary}
STL is a real-time formalism used to describe temporal properties of signals, a finite sequence of states changing over a dense-time domain \cite{stl}. STL grammar is given by
\begin{equation}
   \phi := \top \mid \mu \mid \neg \phi \mid \phi_1 \land \phi_2 \mid \phi_1 U_{I} \phi_2
   \label{eq:stlsyntax}
\end{equation}
where $\top$ is the Boolean True constant, $\mu$ is an atomic preposition of the form $f(\overrightarrow{x}) > 0$ ($\overrightarrow{x}: D \rightarrow \mathbb{R}^l$ is a signal), $\neg$ is negation, $\land$ is conjunction, and $U_I$ is an until temporal operator within $I$, a positive interval.

The logical operators, disjunction/or $\lor$, implies, $\Rightarrow$, and the temporal operators, eventually, F, and always, G, are then derived from the list of expressions in Eq. \ref{eq:stlsyntax} as follows: $\text{F}\phi = \top \mathcal{U}\phi $, $\phi_1 \Rightarrow \phi_2 = \neg \phi \lor \phi_2 $,  $\text{G}\phi = \neg\text{F}\neg\phi$, and $\phi_1 \lor \phi_2 = \neg(\neg\phi_1 \land \neg\phi_2)$.

STL can be interpreted using Boolean semantics (True/False), as well as quantitative semantics (a real-value)\cite{stl}. The quantitative semantics of an STL formula introduces the notion of a robustness value $\rho(\phi, s_t )$ that quantifies the degree to which a formula $\phi$ is violated or satisfied by signal $s_t$, and is given in Table. \ref{tab :STL quantitative semantics}.

\begin{table}[t]
\centering
\renewcommand{\arraystretch}{1.3} 
\caption{STL Quantitative Semantics}
\label{tab :STL quantitative semantics}
\begin{tabularx}{\linewidth}{l l}
\toprule
\textbf{Formula} &  \textbf{Robustness value} \\ 
\hline \hline
$\rho(s_t, >)$ & $\rho_{\text{max}}$ \\
$\rho(s_t, \mu_c)$ & $\mu(x_t) - c$ \\
$\rho(s_t, \neg\phi_1)$ & $-\rho(s_t, \phi_1)$ \\
$\rho(s_t, \phi_1 \land \phi_2)$ & $\min(\rho(s_t, \phi_1), \rho(s_t, \phi_2))$ \\
$\rho(s_t, \phi_1 \lor \phi_2)$ & $\max(\rho(s_t, \phi_1), \rho(s_t, \phi_2))$ \\
$\rho(s_t, \phi_1 \Rightarrow \phi_2)$ & $\max(-\rho(s_t, \phi_1), \rho(s_t, \phi_2))$ \\
$\rho(s_t, \text{F}_{[a,b]}\phi_1)$ & $\max_{t' \in [t+a, t+b]} \rho(s_{t'}, \phi_1)$ \\
$\rho(s_t, \text{G}_{[a,b]}\phi_1)$ & $\min_{t' \in [t+a, t+b]} \rho(s_{t'}, \phi_1)$ \\
$\rho(s_t, \phi_1 \mathcal{U}_{[a,b]}\phi_2)$ & $\max_{t' \in [t+a, t+b]} \Bigl(\min\{\rho(s_{t'}, \phi_2),$ \\
& $\quad\min_{t'' \in [t, t']} \rho(s_{t''}, \phi_1)\}\Bigr)$ \\
\bottomrule

\end{tabularx}
\end{table}


\noindent \textbf{Parametric Signal Temporal Logic (pSTL).} pSTL is an extension of STL where only the structure/template of the STL formula is given, i.e., the STL formula is parameterized and all the time-bounds [$t_1,t_2$] for temporal operators and the constants $\mu$ for inequality predicates are replaced by free parameters \cite{asarin2011exactvalidity}. Parameter valuation $v(p)$ represents a mapping that assigns values to all time and space parameters $p$ of the pSTL. For a given pSTL formula $\phi_p$ with parameters $p$, the valuation of every parameter assignment $v(p)$ results with a corresponding STL formula $\phi_{v(p)}$. In this paper, we only consider unbounded temporal operators, those with time bounds $[0, \infty]$, and will thereby only be concerned with deriving valuations for space parameters.  

\subsection{Reinforcement Learning} \label{ssec:RL}

RL is an optimization problem on a Markov decision process (MDP), a tuple $ M = (S, A, \mathbb{P}, R, \gamma)$ that defines an environment with states $ s \in S$, actions $a \in A$, transition probabilities $ \mathbb{P}(s' | s, a) = \mathbb{P} \{S_{t + 1} = s' | S_t = s, A_t = a \}$, a reward function $ R(s, a) = \mathbb{E}[ R_{t + 1} |  S_t = s, A_t = a ]$, and a discount factor $\gamma \in$ [0,1] prioritizing short term rewards\cite{Sutton1998}. An agent’s behavior is defined by a policy $\pi$ which maps states to a probability distribution over the actions $\pi : S \rightarrow P(A)$, and its objective is to maximize the total discounted return $G_t = \sum_{k = 0}^\infty \gamma^kr_{t + k + 1}$. The state-action value function $Q_{\pi}(s, a) $ is defined as the expected return starting from state $s$, taking action $a$ and thereafter following policy $\pi$
\begin{equation}
    Q_{\pi}(s, a) = \mathbb{E}_{\pi}\begin{bmatrix} G_t | S_t = s, A_t = a \end{bmatrix}
\label{eq: q_func}
\end{equation}

Q-Learning is a foundational value-based algorithm that operates by iteratively approximating the state-action value function $Q(s,a)$ based on the Bellman optimality equation \cite{bellman1952theory}. The update rule is given by  
\begin{equation}
    \resizebox{\hsize}{!}{$ 
Q(s, a) = Q(s, a) + \alpha \left[R(s, a) + \gamma \max(Q(s_{t+1}, a_{t+1})) - Q(s, a)\right]$}
    \label{eq: qupdate}
\end{equation}


For continuous control problems, deep Q-networks (DQN)\cite{lillicrap2019DQN} incorporates neural networks to approximate the Q-value $Q_\theta(s,a) $ parameterized by $\theta$. The optimal parameter $\theta^*$ can be learned using stochastic gradient descent
\begin{equation}\label{eq: thetaopt}
       \theta^* = \arg\min_{\theta} \mathbb{E} \left[(y_t - Q_{\theta}(s_t, a_t))^2\right]
\end{equation}
where  $y_t = r(s_t, a_t) + \gamma \max_a Q_{\theta}(s_{t+1}, a)$ is the temporal difference (TD) target used to stabilize training and maintain a fixed objective over multiple updates.
Policy based methods directly learn a policy $\pi_\theta$ parameterized by $\theta$, that maximizes the expected return from a start state. 
The parameter $\theta$ is updated using gradient ascent
\begin{equation}
\theta_{t+1} \leftarrow \theta_t + \alpha \nabla_\theta J(\pi_\theta)|_{\theta=\theta_t}     
\end{equation}
where $\alpha$ is the learning rate, and $\nabla_\theta J(\pi_\theta)$ is computed following the policy gradient (PG) theorem \cite{Sutton1998}
\begin{equation}
\nabla_\theta J(\pi_\theta) = \mathbb{E}_{\pi_\theta} \left[ Q^{\pi}(s_t, a_t) \nabla_\theta \log \pi_\theta(a_t|s_t) \right]
\end{equation}
TD3 \cite{fujimoto2018TD3}, an algorithm relevant to our proposed approach, is a class of \textit{actor-critic} methods proposed to address the overestimation error caused by deep deterministic policy gradient (DDPG) \cite{lillicrap2019DDPG}. To achieve that, TD3 implements clipped double Q-Learning, delayed policy and target network updates, and target policy smoothing. With these updates, the TD target to which both Q functions regress is given by
\begin{equation}
y_{t} = r(s_t, a_t) + \gamma \min_{i=1,2} Q_{\theta_i'}(s_{t+1}, \pi_{\phi'}(s_{t+1}) + \epsilon)
\end{equation}
where $Q_{1,2}$ are the critic networks, $\pi$ is the actor network, $\theta_i'$ and $\phi'$ are the target critic and target actor network parameters, respectively, and $i = 1, 2$ represents the $i$-th target critic networks, and $\epsilon$ is the clipped Gaussian noise.

\noindent \textbf{Constrained RL.} It is a branch of RL that is concerned with maximizing reward while also satisfying environmental safety constraints. Safe RL is modeled as a constrained MDP (CMDP)\cite{altman-constrainedMDP}, which is an extension of the standard MDP with an additional constraint set $\mathcal{C}$. The optimal policy in constrained RL is expressed as 
\begin{equation}
    \pi^* = \arg\max_{\pi \in \Pi_c}  J^\mathcal{R}(\pi)
\end{equation}
where $J^\mathcal{R}(\pi)$ is the objective function and $\Pi_c$ is a set of constraint satisfying policies.



\subsection{Bayesian Optimization} \label{ssec:BO}

BO is a powerful strategy for the optimization of black-box functions that are intractable to analyze and are often non-convex, nonlinear, and computationally expensive to evaluate \cite{frazier2018tutorial}. It has been widely applied across multiple fields, such as hyperparameter tuning in machine learning models \cite{victoria2021automatic}, control and planning \cite{baheri2020waypoint,baheri2017real,baheri2019combined}, robotics \cite{calandra2016bayesian}, and materials design \cite{zhang2020bayesian}. BO offers a principled technique to direct a search of the global optimum of an objective function by building a probabilistic model of the objective function, called the surrogate function, that is then searched efficiently guided by an acquisition function. GPs are nonparametric models employed in BO to impose a prior over the objective function. GP is used to maintain a belief over the design space simultaneously modeling the predicted mean \( \mu(p) \) and the epistemic uncertainty \( \sigma(p) \) at any parameter set \( p \) in the input space. GPs are defined by their mean function \( \mu(\mathbf{p}) \), which is initially assumed to be 0, and covariance function \( k(p,p') \)
\begin{equation}
    f(p) \sim \mathcal{GP}(\mu(p), k(p,p'))
\end{equation}
The covariance function \( k(p,p') \) is also called the \enquote{kernel}, and is often given by a squared exponential function
\begin{equation}
    k(p, p') = \exp\left( -\|p - p'\|^2 \right)
\end{equation}

For any new set of parameters \( p_* \) for the pSTL, the GP model provides a predictive distribution with mean and variance given by \cite{BayesianOptimization}
\begin{equation}
 \begin{aligned}
\mu(p_*|p) = \mu(p_*) + \mathbf{K}_*^T \mathbf{K}^{-1} (\mathbf{y} - \mu(p)) \\
\sigma^2(p_*|p) = \mathbf{K}_{**}  - \mathbf{K}_*^T \mathbf{K}^{-1} \mathbf{K}_*
\end{aligned}  
\label{eq:gpmodel}
\end{equation}
where \( \mathbf{K} = k(p, p) \), \( \mathbf{K}_* = k(p, p_*) \), and \( \mathbf{K}_{**} = k(p_*, p_*) \).


Acquisition functions guide how the parameter space is explored by observing the predicted mean and variance of a sample parameter set from the GP model, given in Eq. \ref{eq:gpmodel}. We use the expected improvement (EI) as the acquisition function. EI accounts for the size of improvement over the current best observation when choosing the next candidate parameter set. The utility of EI lies in its ability to explicitly encode a trade-off between pursuing regions of high uncertainty (exploration) and regions with a potential for high objective function values (exploitation) by quantifying the expected amount of improvement.
The EI for a parameter set \( p \), given the current observations $\mathcal{D}$, is defined as follows \cite{BayesianOptimization}
\begin{equation}
EI(p) = \mathbb{E}[\max(0, f_{\min}(p) - f(p)) | p, \mathcal{D}]
\end{equation}
where \( f_{\min} \) is the minimum value observed so far.
Through iterative implementations of the acquisition function, the GP model refines its predictions, steering the optimization process towards the global optimum of the objective function.

\section{Problem Statement and Formulation}\label{sec:problem statement}
We consider the problem of safe RL policy synthesis in an environment where safety constraints are unknown \textit{a priori}. Our ultimate objective is to concurrently learn accurate parameters of the pSTL specification that define the environmental constraints and an optimal policy such that the policy adheres to the learned STL safety constraint while achieving high returns. 

We initialize the problem with a small initial labeled datasets, safe trajectories $D_s$ and unsafe trajectories $D_{us}$, and a pSTL safety specification template $\phi_p$. Within the initial labeled datasets $D_{s}$ and $D_{us}$, we find safe traces $x_{s}$ and unsafe traces $x_{us}$, respectively, which are initially manually selected by the human expert from historical safe and unsafe runs within the given environment. The small size requirement of these initial datasets relieves the difficulties of acquiring a large pool of expressive historical datasets that is required for learning of accurate pSTL parameters from data, especially in safety-critical environments. Our approach, instead, takes on a data-efficient strategy that starts with a small initial dataset and iteratively adds to it additional expert-labeled data as necessary to strategically acquire high-quality parameter estimates using the smallest applicable volume of labeled data. 

The pSTL parameter learning process takes on a supervised learning approach that requires obtaining parameter valuations for the pSTL, such that the synthesized STL formula is satisfied by safe trajectories and  is not satisfied otherwise. The goal is to learn sufficiently accurate set of parameters of the pSTL with smallest viable dataset size. The parameter synthesis problem focuses on solving a minimization problem for a black-box objective function with the aim of finding optimal parameters of the pSTL specification using a labeled dataset.

The policy learning step requires solving the optimization problem in CMDPs expressed as 
\begin{equation}
    \begin{split} \max_{\pi_\theta \in \Pi_C} &J^\mathcal{R}(\pi_\theta) \\
\text {s.t.}~~& J^{\mathcal{C}}(\pi_\theta) \leq d\end{split}
\label{eq:cmdp}
\end{equation}
where $J^\mathcal{R}$ is a reward-based objective function, $J^{\mathcal{C}}$ is a cost-based constraint function, $\Pi_C$ denotes a feasible set of constraint satisfying policies, and $d$ is the threshold for safety. Within our framework, we formulate the cost objective function $J^\mathcal{C}$ by finite-horizon, undiscounted expected cumulative costs  
\begin{equation}
  J^{\mathcal{C}} = \mathbb{E}_{\tau \sim \pi_{\theta}} [ \sum_{t=0}^{T} c_t] 
\end{equation}
where $c_t$ is the cost at time-step $t$ and is computed using the STL safety constraint. 
The learned policy is considered optimal if  achieves its performance objectives while also generating rollout traces that demonstrate constraint-abiding behavior. \emph{Rollouts} are sequences of actions executed by an agent from a specific state under the learned policy, thus serve as direct indicators of the policy's safety.

\section{Methodology}\label{sec:Methodology}

In our proposed framework, the learning of a safe RL policy in an environment with unknown safety constraint parameters is separated into two components: optimization of the parameters for the given pSTL safety constraint using labeled data, and safe RL policy optimization with logical constraints. These two components are integrated through the assistance of a human expert, who contributes by labeling rollout traces derived from an RL policy. The labeling process involves the human expert designating each rollout trace as \enquote{safe} or \enquote{unsafe} based on whether each trace adheres to or violates safety constraints in an environment. In other words, a trajectory is labeled \enquote{safe} if and only if all the states in the trajectory are safe as deemed by the human expert, and labeled \enquote{unsafe} otherwise. This labeling process is crucial, as it yields the labeled dataset required for the iterative refinement of the pSTL parameters. 

We frame this concurrent learning problem as a bilevel optimization, an optimization approach that contains two levels of optimization tasks where one optimization task, the lower level, is nested within the other, the upper level\cite{BLO}. These two levels of optimization each address one each of the two learning components in our framework: the upper level is dedicated to the pSTL parameter synthesis while the lower level is dedicated to the constrained RL policy optimization. The mathematical representation of this bilevel optimization task is given by
\begin{equation}
  \begin{aligned}
    \quad &\arg\min_{p}  f(\phi_{v(p)}, \pi^*(\phi_{v(p)})) ,\\
    \textrm{s.t.} \quad &  \pi_{\theta}^*(\phi_{v(p)})  \in \arg\max_{\pi_\theta \in \Pi_c}  J^\mathcal{R}(\pi_\theta(\phi_{v(p)}) )
\end{aligned}  
\label{eq:BLO}
\end{equation}
where $f$ is the upper-level objective function with optimization variable $p$, which is a set of parameter values to the pSTL $\phi_p$, and $\pi$ is the lower-level optimization objective with optimization variable $\theta$. $\pi^*(\phi_{v(p)})$ represents the optimal policy under the given STL constraint $\phi_{v(p)}$, and $\phi_{v(p^*)}$ represents the complete STL after the valuation of pSTL $\phi_{p}$ with optimal parameters $p^*$. In Eq. \ref{eq:BLO}, the upper-level objective $f$ depends on both the pSTL parameters $p$, and the solution $\pi^*(\phi_{v(p)})$ of the lower-level objective. A schematic of our framework is illustrated in Fig. \ref{fig:framework}. 

\begin{figure*}[!h]
\centering
    \includegraphics[trim={1.5cm 6.5cm 1.2cm 10.8cm},clip,width=\linewidth]{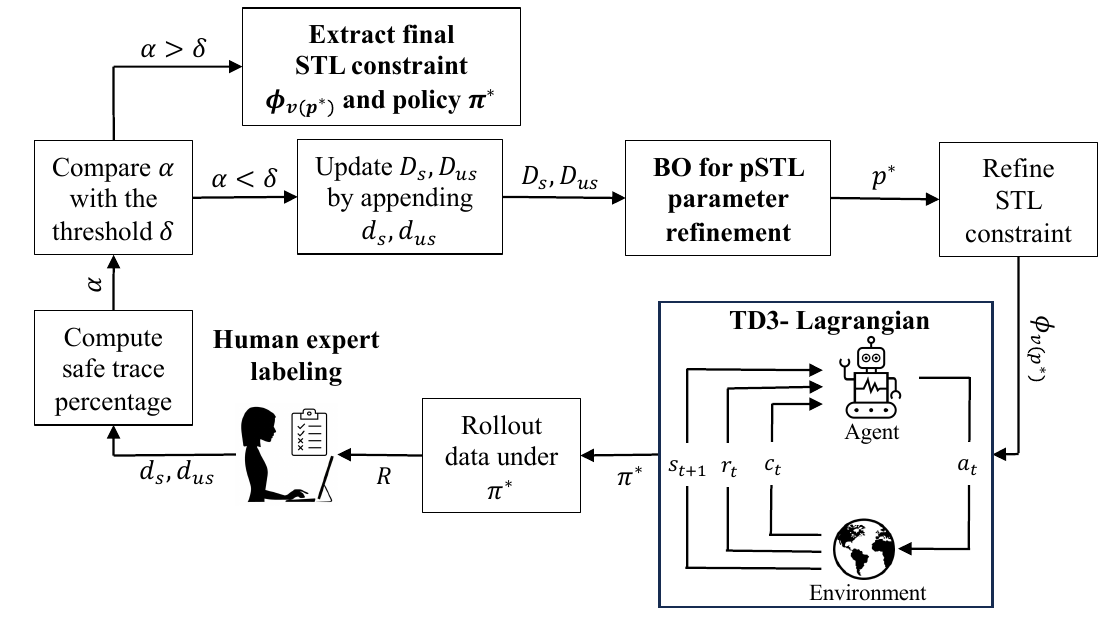}
    \caption{Schematic representation of the integrated framework for concurrently learning STL constraint parameters and optimal policies. The framework applies BO for STL parameter mining, TD3-Lagrangian for policy learning, and incorporates human expert for labeling rollout traces to be used in refining the learned constraint parameters and policy. Once the percentage of safe traces in a rollout dataset $\alpha$ is higher than the threshold value $\delta$, convergence is achieved, and the final policy and STL constraint are extracted.}
\label{fig:framework}
\end{figure*}

\subsection{STL Constraint Parameter Learning}\label{ssec:STLLearning}
The upper-level of the bilevel optimization framework, a BO process, is designed to obtain the optimal parameters $p^*$ of a given pSTL formula $\phi_p$ (an STL formula template) through the minimization of an objective function $f$. The parameter learning process initiates with the pSTL formula $\phi_p$ and the two initial safe and unsafe datasets $D_s, D_{us}$. Using these labeled datasets, BO is carried out to learn the best parameter configuration for pSTL $\phi_p$ such that the final STL best classifies between $D_s$ and $D_{us}$ in terms of robustness degree. The rationale behind designing the objective function for learning the optimal parameters of the pSTL specification $\phi_p$ is as follows: if a candidate STL $\phi_{v(p)}$ represents the true environmental constraints, any trace labeled \enquote{safe} by the human expert $x_s$ should have a positive robustness value and any trace labeled \enquote{unsafe} $x_{us}$ should have a negative robustness value with respect to $\phi_{v(p)}$. Under this consideration, the objective function used for pSTL parameter optimization $f$ is mathematically defined by
\begin{equation}
f(\phi_{v(p)}) =  \frac{1}{2}\left(\frac{N_{\rho(\phi_{v(p)})\textsuperscript{-}\mid x_{s}}}{N_{x_{s}}} +\frac{ N_{\rho(\phi_{v(p)})\textsuperscript{+}\mid x_{us}}}{N_{x_{us}}}\right) 
\label{eqn:objfunction}
\end{equation}
where $\phi_{v(p)}$ is the STL formula obtained by the parameter valuation $v(p)$ of pSTL $\phi_p$, $x_s$ and $x_{us}$ are traces sampled from the datasets containing safe and unsafe traces, respectively. $N_{x_s}$ and $N_{x_{us}}$ are the total number of safe and unsafe traces within their respective datasets. The first term within the parenthesis in Eq. \ref{eqn:objfunction} represents ratio of safe traces $x_s$ with a negative robustness value $\rho^-$ (false negative rate), and the second term represents ratio unsafe traces $x_{us}$ with a positive robustness value $\rho^+$ (false positive rate) with respect to $\phi_{v(p)}$. This essentially computes the balanced misclassification rate, derived from the complement of the balanced accuracy score \cite{balancedacc}, a metric that computes classification accuracy in datasets with imbalanced distribution between classes. 

Upon convergence, this optimization process will identify the optimal set of parameters $p^*$ for the  pSTL that minimize the objective function, $f$ in Eq. \ref{eqn:objfunction}, yielding the final STL safety constraint $\phi_{v(p^*)}$, which we denote $\phi_{cost}$. $\phi_{cost}$, is of type $\text{G}(\lnot(\psi))$ where $\psi$ characterizes unsafe behavior, and $\phi_{cost}$ conveys \textit{\enquote{always-not-unsafe}} i.e. $G (\lnot (\psi))$, signifying that it universally opposes the occurrence of $\psi$.

\subsection{Policy Learning}\label{ssec: policylearning}
The lower-level of the bilevel optimization framework consists of a  logically-constrained, safe RL policy optimization. This phase follows the process of pSTL parameter optimization, detailed in Section \ref{ssec:STLLearning}, and uses the STL generated therein as its input. For this stage, we solve the optimization problem for CMDP introduced in Eq. \ref{eq:cmdp} by utilizing the Lagrangian-variant of the twin delayed deep deterministic policy gradient (TD3) algorithm, TD3-Lagrangian. The background for the TD3 algorithm is given in Section \ref{ssec:RL}, and throughout this section, we provide an overview of Lagrangian methods, later discussing the development of the TD3-Lagrangian algorithm.

Lagrange multiplier method is used to transform a constrained optimization problem into an equivalent unconstrained optimization problem through Lagrangian relaxation procedure that introduces adaptive penalty coefficients to enforce constraints \cite{bertsekas2014constrainedRLLagrange}. Using this method, Eq. \ref{eq:cmdp} is transformed into the equivalent unconstrained min-max optimization problem 
\begin{equation}
    \max_\theta \min_{\lambda \geq 0} \mathcal{L}(\theta, \lambda) =J^\mathcal{R}(\pi_{\theta}) - \lambda ( J^{\mathcal{C}}(\pi_{\theta}) - d)
    \label{eq:lagrangian}
\end{equation}
where $\lambda$ is the Lagrange penalty coefficient, $J^\mathcal{R}$ is the reward objective function, $J^\mathcal{C}$ is the constraint objective function, and $d$ is the maximum allowable cumulative cost. Eq. \ref{eq:lagrangian} is then solved by gradient ascent on $\theta$ and descent on $\lambda$ to result with the optimal values $\theta^*$ and $\lambda^*$. 

An adaptation of the Lagrange multiplier method to the TD3 algorithm is given in \cite{ji2023omnisafe}, deriving TD3-Lagrangian. TD3-Lagrangian incorporates an additional cost critic network to the original TD3 architecture to estimate the cost value function $Q^\mathcal{C}$, and alters the loss function to incorporate a constraint satisfaction component using a Lagrangian multiplier
\begin{equation}
L = -Q^V (\pi_{\theta}, s) + \lambda Q^\mathcal{C} (\pi_{\theta}, s)
\label{eq: LossFunc}
\end{equation}
where $Q^V$ is the minimum value of the two reward critic network outputs, $Q^C$ is the value of cost critic network, and $\pi$ is the policy network. The penalty coefficient $\lambda$ is  updated by minimizing the penalty loss $\lambda ^{'} = \lambda + \eta (J^\mathcal{C}(\pi_{\theta}) - d)$, where $\eta$ is the learning rate. When $J^\mathcal{C}$ exceeds the constraint threshold $d$, $\lambda$ is increased to prioritize cost minimization.

In our approach, we propose a novel modification to the TD3-Lagrangian architecture. While retaining the classical definition of the reward function for each environment, we  redefine the cost function logically, using an STL specification $\phi_{cost}$. As stated prior, $\phi_{cost}$ is the STL safety specification derived through the process outlined in Section \ref{ssec:STLLearning}. Using this STL as the safety constraint, we compute the cost at each step $c(s_t, a_t)$, using the quantitative semantics of STL given in Section \ref{ssec:STLPreliminary}, as follows
\begin{equation}
    c(s_t, a_t) = 
\begin{cases}
   1 ,& \text{if }\rho(\phi_{cost},s_t) < 0\\
   0,  & \text{if } \rho(\phi_{cost},s_t) \geq 0
\end{cases}
\label{eqn:costfun}
\end{equation}
where $\rho (\phi_{cost},s_t)$ is the robustness value of the current state $s_t$ with respect to the STL $\phi_{cost}$. This equation is interpreted as follows. The cost $c(s_t, a_t)$ is assigned to 1 if $\rho (\phi_{cost},s_t) < 0$, indicating the safety constraint has been violated at state $s_t$, and to 0 otherwise. We use this STL robustness-based cost values to compute $J^\mathcal{C}$, which is then used in the policy optimization process by minimizing the loss function in Eq. \ref{eq: LossFunc}.


Once we have identified an optimal policy $\pi^*$, we generate a dataset of rollout traces $R$ by executing the learned policy, which we then pass on $R$ to the human expert for labeling. 

\subsection{Human Feedback Mechanism}\label{ssec:humanfeedbackmech}
A crucial role is played by a human expert who provides labels to the rollout traces generated through the execution of the RL policy optimized through the process outlined in Section \ref{ssec: policylearning}. This labelling process is essential to our framework in gradually and iteratively refining the parameter assignment for the pSTL. The quality of the parameters assigned is reliant on the volume of labeled dataset, and while the richness of data facilitates this process, extensive human labeling effort to amass large datasets is impractical. Hence, our strategy focuses on attaining sufficiently accurate pSTL parameters with the minimal necessary data. This is achieved by the human expert labeling only a small number of traces at each iteration, which are then incrementally added to the existing dataset of labeled data from previous iterations. The phased acquisition of data across different iterations of RL policy rollouts ensures that each new rollout data set is attained from a unique policy and contributes unique and essential information to the learning process, enhancing the overall quality and diversity of the dataset used for pSTL parameter learning.

During our experiments, we have implemented an automated process for labeling the rollout traces, which involves of computing the robustness value of each trace within the rollout set with respect to the true STL safety constraint $\psi$. The human labeling process is given using the satisfaction relation $\models$ between a trace from the rollout dataset $x$ and an STL formula $\psi$ as follows

\begin{equation}
    L(x) = \begin{cases}
   1 ,& \text{if } x \models \psi\\
   0, & \text{if } x \not\models \psi\\
\end{cases}
\label{eqn:labeling}
\end{equation}
where $L(x)$ is the label assigned to trace $x$ sampled from the rollout dataset, $\psi = G_{[0:T]}( \lnot( \phi_{true} ))$ is the general template we use for any STL safety constraint in which $\phi_{true}$ is the environment-specific STL formula exhibiting unsafe behaviour. Eq. \ref{eqn:labeling} states that a label of 1 is assigned if a trace from the rollout dataset $x$ satisfies ($\models$) the true STL safety constraint $\psi$, i.e. $\rho(\psi,x) > 0 $, and a label of 0 is assigned otherwise. 

It is important to note that the use of the true STL safety constraint for automation purposes is not to be confused with the algorithm having the knowledge of this true STL environmental constraint beforehand. In real-world applications, as per the basis of our problem statement, the actual safety constraint remains unknown to the algorithm and is only used for rapid and efficient experimentation of our framework across various case studies, to which we do not have an actual human expert in those areas to label the traces. Therefore, due to the unavailability of the True environmental constraint in practical implementations of our framework, the involvement of the human expert in manually labeling the traces is integral to our framework. After traces are labeled, those identified as safe by the human expert are allocated in the safe dataset $d_s$, whereas those labeled as unsafe are allocated to the unsafe dataset $d_{us}$. Finally, the percentage of safe rollout traces $\alpha$ within the rollout dataset $R$ is computed by
\begin{equation}
    \alpha = \left(\frac{N_{d_s}}{N_{d_s} + N_{d_{us}}} \right) 
    \label{eq:percentage safe}
\end{equation}
where $\alpha$ is the percentage of safe traces, $N_{d_s}$ and $N_{d_{us}}$ are the respective number of traces in the safe dataset $d_{s}$ and unsafe dataset $d_{us}$. The sum $N_{d_s} + N_{d_{us}}$ is equal to the total number of traces in the rollout dataset $N_R$.


\begin{algorithm}[t]
\caption{Joint Learning of Policy with Constraints}
\label{alg:jlf}

\textbf{Input:} $D_s, D_{us}$, $\phi_{p}$,  $n_R$, $n_s$, $\delta$\\
\textbf{Output:} $p^*$, $\pi^*$

\begin{algorithmic}[1] 
\STATE Compute $\alpha$ from initial datasets $D_s, D_{us}$ using Eq. \ref{eq:percentage safe}
\WHILE{$\alpha < \delta$}
    \STATE $p^{*} \leftarrow$ Employ BO to find optimal parameters of $\phi_{p}$ using $D_s, D_{us}$ by minimizing Eq. \ref{eqn:objfunction}
    \STATE $\phi_{v(p^{*})} \leftarrow $ pSTL valuation using optimal parameters $p^{*}$
    \FOR{$i \gets 1$ to $n_s$}
        \STATE $\pi^* \leftarrow$ Employ TD3-Lagrangian to optimize safe policy under STL constraint $\phi_{v(p^{*})}$
    \ENDFOR
    \STATE $R \leftarrow $ Generate $n_R$ rollout traces under $\pi^*$ and generate dataset of the traces 
    \FOR{ Trace $\in R$}
        \STATE Human expert provides a \enquote{safe}/\enquote{unsafe} label to Trace \\
        \IF{Trace is labeled safe}
            \STATE Store in safe dataset $d_s$
        \ELSIF{Trace is labeled unsafe}
            \STATE Store in unsafe dataset $d_{us}$
        \ENDIF
    \ENDFOR
    
    \vspace{2pt}
    \STATE $\alpha \leftarrow$ Compute percentage of safe traces in $R$ from datasets $d_s, d_{us}$ following Eq. \ref{eq:percentage safe}
    \vspace{1pt}
    \STATE \textbf{Append} $d_s, d_{us}$ \textbf{to} $D_s, D_{us} \leftarrow $ Extend initial dataset with new labeled rollout traces
\ENDWHILE
\end{algorithmic}
\end{algorithm}

Our framework is outlined in Algorithm \ref{alg:jlf}. The algorithm requires as input initial  datasets, $D_s$ and $D_{us}$, populated with safe traces and unsafe traces, respectively, and the pSTL specification $\phi_p$, along with the number of rollout traces at each iteration $n_R$, the total RL training steps $n_s$, and the user-specified threshold for the minimum satisfactory percentage of safe traces within a rollout dataset $\delta$. The algorithm initiates by using BO to optimize parameters to the pSTL from the initial dataset, then optimizes a policy using TD3-Lagrangian algorithm constrained by the learned STL. It then proceeds to generate rollout traces from this policy, which are subsequently labeled by a human expert. A key metric, $\alpha$, is then calculated representing the percentage of \enquote{safe} rollout traces as labeled  by the human expert amongst the entire rollout dataset and is compared to $\delta$ at every iteration. 

This iterative process is repeated until convergence,  which is achieved when $\alpha$ is greater than the user-specified, minimum threshold for the percentage of safe rollout traces $\delta$ i.e. $\alpha > \delta$. If the convergence criteria has not been met, $d_s$ and $d_{us}$ are appended to the initial datasets $D_s$ and $D_{us}$ respectively, to then serve as inputs for the next iteration to generate new and refined STL parameters. If the convergence criteria has been met, the outputs of the algorithm, an STL with optimal parameter values $\phi_{v(p^*)}$, and the optimal policy $\pi^*$ with respect to the learned STL, are extracted. 

\section{Case Studies}\label{sec:case studies}
We implement our concurrent learning framework across a series of case studies described throughout this section. All case studies were performed on Safety-Gymnasium environments \cite{ji2023safetygymnasium}, a safe RL benchmark that comprises of several safety-critical tasks in continuous control environments where agents and tasks are inherited from safety-gym\cite{Achiam2019BenchmarkingSE} and MuJoCo physics simulator \cite{mujoco2012}\footnote{Our implementation for this research is available at our GitHub repository: \url{https://github.com/SAILRIT/Concurrent-Learning-of-Control-Policy-and-Unknown-Constraints-in-Reinforcement-Learning.git}}.

\subsection{Case Study 1: Safe Navigation - Circle}\label{ssec:cs1}
\begin{wrapfigure}[15]{r}{0.45\linewidth} 
  \centering
  \includegraphics[width=\linewidth]{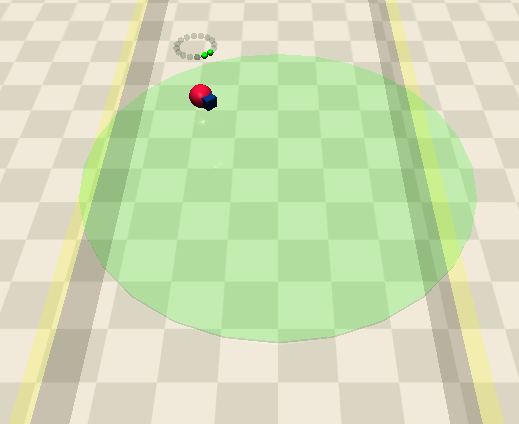} 
  \caption{Circular navigation environment with 2 boundaries in the $x$ direction (in yellow) and the safe navigation area (in green).}
  \label{fig:spc1}
\end{wrapfigure}

Safety navigation-circle offers a scenario in which an agent is situated randomly within a given $x$ and $y$ bounds at the start of an episode. The objective of the agent is to move in a circular motion within the circle area, while also attempting to stay at the outermost circumference of the circle. In doing so, the agent must also avoid going outside safety boundaries that intersect with the circle area as depicted in Fig. \ref{fig:spc1}. We use level 1 of this environment as given by \cite{ji2023safetygymnasium}, which consists of 2 boundaries, situated on the left and right side of the of the center, respectively, and the $point$ agent \cite{Achiam2019BenchmarkingSE}, a simple robot constrained to a 2D plane with two actuators, one for rotation and the other for forward/backward movement. The reward function for this environment is given as \cite{ji2023safetygymnasium}
\begin{equation}
r_t = \frac{1}{1 + |r_{a} - r_{c}|} \cdot \frac{(-u \cdot y + v \cdot x)}{r_{a}}
\label{eq:rewardspc}
\end{equation}
where $r_t$ is the current time-step reward, $u, v$, are the $x-y$ axis velocity components of the agent, $x ,y$ are the $x-y$ axis coordinates of the agent, $r_{a}$ is the Euclidean distance of the agent from the origin, $r_{c}$ is the radius of the circle geometry. Intuitively, the agent moves as far out as it can in outermost circumference of the circle, and the faster the speed, the higher the reward. The predefined pSTL that represents the STL safety constraint template of this environment is given by
\begin{equation}
    \phi_{\text{cost}} = G\left(\lnot(\left(x_{a} < x_{\mathcal{T}^-}) \vee (x_{a} > x_{\mathcal{T}^+})\right)\right)
\label{eq:pSTLspc}
\end{equation}
where $x_{a}$ represents the agent’s $x$ position, $x_{\mathcal{T}^+}$ and $x_{\mathcal{T}^-}$ represent the x threshold locations where the boundaries in the positive and negative x directions, respectively, are located, measured from the center.

The pSTL specification in Eq. \ref{eq:pSTLspc}, intuitively describes that the agent’s $x$ location should never move past the boundaries in either direction of the center. The (initially unknown) safety constraint parameters for this environment are the threshold values $x_{\mathcal{T}^+}$ and $x_{\mathcal{T}^-}$, which provides us with two learning parameters for this pSTL to obtain an STL safety specification.
\subsection{Case Study 2: Safe Navigation - Goal} \label{ssec:cs2}

\begin{wrapfigure}[14]{r}{0.45\linewidth} 
  \centering
  \includegraphics[width=\linewidth]{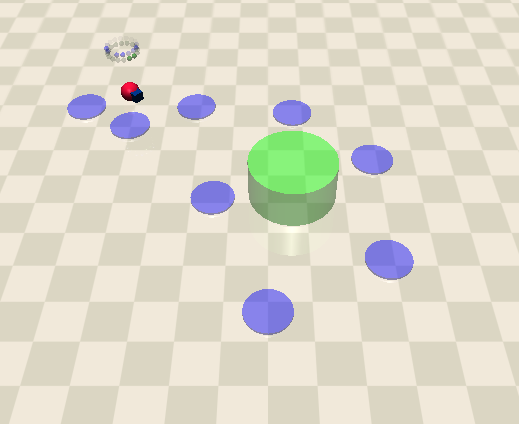} 
  \caption{Goal navigation environment with eight hazards (in blue), and one goal location (in green).}
  \label{fig:spg1}
\end{wrapfigure} 

Safe Navigation-Goal is another environment introduced in \cite{ji2023safetygymnasium} that offers a scenario in which an agent is randomly positioned at the start of an episode, with the objective of navigating to a designated goal location within the environment while circumventing circular hazard locations. Upon reaching the designated target location, this location is reassigned randomly to a new goal location and the agent continues to navigate towards the updated target. This process continues until the maximum episode steps is reached. We implement level 1 of this environment, which comprises of 8 hazard locations and one goal location. Similarly to \ref{ssec:cs1}, we use the \enquote{$point$} agent within this environment. A snapshot of the environment is shown in Fig. \ref{fig:spg1}. The reward function for this environment is defined as \cite{ji2023safetygymnasium}
\begin{equation}
r_t = (d_{\text{t-1}} - d_{\text{t}}).\beta   
\label{eq: rewardgoal}
\end{equation}
where $r_t$ represents the reward at the current time step, $d_{\text{t-1}}$ and $ d_{\text{t}}$  represent Euclidean distances between the agent $a$ and the goal $g$ at the previous time step $t-1$ and the current time step $t$, respectively, and $\beta$ is a discount factor. When $d_{\text{t-1}} > d_{\text{t}}$, it indicates that the agent is moving closer to the goal, and $r_t > 0$ as a result and vice versa.

The pSTL safety constraint that is provided for this environment is given by
\begin{equation}
     \phi_{\text{cost}} =   G\Biggl(\neg \biggl(\bigvee_{i=1}^{8}\left(\sqrt{(x_{a}-x_{h,i})^2+(y_{a}-y_{h,i})^2} < r_{h}\right)\biggr)\Biggr)
\label{eq:pSTLspg}
\end{equation}
%
where $i=1,2,\ldots,8$ represents each of the 8 hazards, $x_{h,i}$ and $y_{h,i}$ are the $x$ and $y$ coordinates of hazard $i$'s centroid, respectively, $x_{a}$ and $y_{a}$ are the agent's current $x$ and $y$ location, respectively, and $r_{h}$ represents the radius of the hazards.

The pSTL expression provided in Eq. \ref{eq:pSTLspg} can be interpreted as follows. The Euclidean distance between the agent and any of the eight hazards should never be less than the hazard's radius. The initially unknown safety constraint parameters for this environment are the $x$ and $y$ coordinates of the centroids of the hazards: $x_{h,i}$ and $y_{h,i}$, where $i = 1,2,\ldots,8$ represents each of the 8 hazards, adding up to 16 unknown parameters to learn. 

\subsection{Case Study 3: Safe Velocity - Half Cheetah}
\begin{wrapfigure}[12]{r}{0.45\linewidth} 
  \centering
  \includegraphics[width=\linewidth]{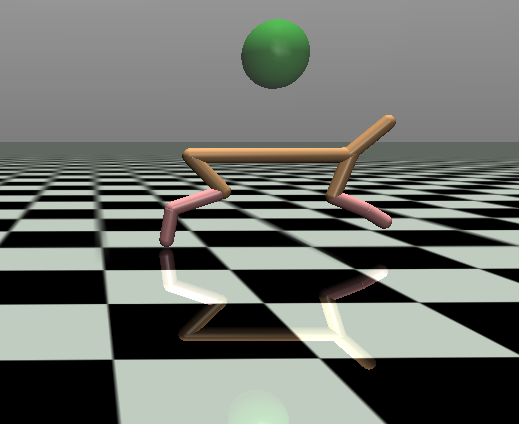} 
  \caption{Safe velocity test environment with the half cheetah agent.}
  \label{fig:shcv}
\end{wrapfigure}
The Half Cheetah environment provided by \cite{brockman2016openai, mujoco2012} features a two-dimensional, half-body of a cheetah consisting of 9 body parts and 8 joints connecting them as shown in \ref{fig:shcv}. The state space of this environment includes the positions, angles, velocities, and angular velocities of the cheetah's joints and segments whereas the action space is defined by the torques applied to these joints. The primary objective for this agent is to apply torque on the joints to make the cheetah run in the forward direction to achieve maximum  speed, $u$. There is, however, a control cost penalty applied to restrict the agent from taking too large actions. Overall, the reward allocated to the agent is based on the forward movement and control cost penalty, which is calculated as the weighted sum of the squares of the actions (torques)
\begin{equation}
    r_t = \left(w_{f} \cdot \frac{x_{t-1} - x_{t}}{d_t}\right) - \left(w_{c} \cdot \sum(a_{t}^2)\right)
    \label{eq:hcvreward}
\end{equation}
where $w_f$ if the forward reward weight, $x_{t-1}$ and $x_{t}$ are the x-coordinates of the agent before and after applying action $a_t$, respectively,  $d_t$ is the time between actions, and $w_c$ is the control cost weight. 
A refined adaptation of the Half Cheetah environment is detailed in \cite{ji2023safetygymnasium}, introducing an additional constraint on the agent's maximum allowable x-velocity. We use this adaptation for our experimentation on this environment with the pSTL safety constraint given by
\begin{equation}
    \phi_{\text{cost}} = G\left(\lnot( u_{a} > u_{max} \right))
\label{eq:pSTLshcv}
\end{equation}
where $u_{a}$ is the agent's x-velocity, and $u_{max}$ is the the maximum allowable (safe) x-velocity for the agent. The pSTL given in Eq. \ref{eq:pSTLshcv} provides one  parameter to be learned using our framework, $u_{max}$.

We evaluate key performance metrics of out approach through the two primary tasks: 1) optimization of safe policies, and 2) synthesis of pSTL parameters. In regards to safe policy optimization, we first show convergence during policy optimization for all case-studies alongside a comparative analysis of cumulative rewards and costs per episode at the end of training against established baselines. Furthermore, we rigorously evaluate the policies by examining the safety of rollout traces generated under each policy. Pertaining to the synthesis of pSTL formula parameters, our evaluation focuses on comparing the learned parameters against the True environmental safety parameters, which are unknown \textit{a priori} to the algorithm. We also evaluate the classification accuracy of the learned STL safety constraints with respect to labeled data. We compare our algorithm with:

\begin{itemize}
    \item \textbf{Baseline 1} : Unconstrained RL policy optimization in an environment in which safety constraints are unknown, which is a condition we were faced with prior to implementing our framework,  and 
    \item \textbf{Baseline 2}:  Constrained RL policy optimization in an environment with known STL safety constraint. 
\end{itemize}

The rationale behind the selection of these two baseline approaches to compare to ours is as follows: baseline 1, involving unconstrained policy optimization, underscores the criticality of clearly defining safety constraints and elucidates the safety risks associated with deploying algorithms trained in the absence of appropriate safety constraints. In contrast, baseline 2 represents an optimal scenario wherein all environmental safety constraint parameters are known a priori, facilitating a comparative analysis to gauge the proximity of our framework's results to this ideal benchmark.

We chose to consider $\alpha$ as the principal convergence metric because it evaluates success in both the upper and lower level optimization problems, i.e. it serves as a qualitative indicator of the effectiveness of the learned STL safety specification in guiding the cost assignment during policy optimization as judged by the human expert and the ability of the RL algorithm to generate a policy that adheres to STL safety constraints.

We implemented our proposed algorithm and the two baseline methods to optimize a policy within the each of the specified case studies. The experiments were conducted under consistent environmental settings, with the primary distinction being in the computation of costs for each method at each step. The training parameters used are given in Table. \ref{tab4}. 
\begin{table}[t]
\caption{Training Hyper-parameters}
\label{tab4}
\centering
\begin{tabular}{@{}c*{6}{c}}
\toprule 
\textbf{Hyper-parameter} & \textbf{Value} \\
\midrule \midrule
Actor learning rate & $5\cdot 10^{-6}$   \\
Critic learning rate  &  $10^{-3}$   \\ 
Discount factor & $0.9$  \\ 
Batch size & $256$  \\ 
Policy update delay & $2$  \\ 
Exploration noise & 0.1\\
Policy noise & 0.2\\
Policy noise clip & 0.5\\
Actor/Critic activation function & ReLU\\
Total steps (case-study 1,2,3) & $10^6, 1.5\cdot 10^6, 10^6$ \\ 
Steps per epoch (case-study 1,2,3) & $5\cdot10^2, 10^3, 10^3$  \\ 
Cost limit & $0.0$  \\ 
$\lambda$ learning rate & $5 \cdot 10^{-7}$ \\
$\lambda$ optimizer & Adam\\
\bottomrule
\end{tabular}
\label{tab: hyperparams}
\end{table}

Specifically for our algorithm, the cost assignment during policy optimization is based on a learned STL safety constraint following the process detailed in Section \ref{ssec: policylearning}. This contrasts with baseline 2, where the cost assignment stems directly from the actual STL safety constraint, a value which, in practice, is unknown. Finally, baseline 1, an unconstrained optimization approach, does not incorporate cost considerations due to the absence of known safety constraints in the context. For experimentation within our framework, the convergence threshold, $\delta$, was set to 90\% for case studies 1 and 3, and to 75\% for case study 2, indicating the algorithm terminates once the rollout trace from a policy attains the specified percentage of safe traces. These numbers were decided based on the complexity of the environment, specifically in relation to the quantity of safety parameters required for learning.  Convergence was attained when implementing our framework, on average, after 9 iterations for case study 1, 17 iterations for case study 2, and 6 iterations for case study 3. The initial dataset of labeled data contained 10 safe traces and 10 unsafe traces in $D_s$ and $D_{us}$, respectively, and at each iteration, 50 rollout samples are provided to the human expert for labeling. Through experimentation, we determined that labeling 50 rollout traces per iteration effectively manages the human expert's workload while ensuring that the learning process remains unimpeded. The safety threshold (cost limit) in the constrained optimization settings is set to 0, indicating that no violations of safety constraints are permissible at any point in the trajectory for it to be deemed safe.

\section{Results and Discussion}{\label{sec: results and discussion}}
We first delve into the results obtained from the BO process employed for pSTL parameter learning. To illustrate these outcomes, we have included the learning curve of the BO, which is depicted in Fig. \ref{fig:BO} and we present the learned STL safety specification, valuated with the optimal parameters obtained through the BO process, alongside the true STL specifications for easy comparison in Table. \ref{tab:Learned and True STLs}. Using the Learned STLs given in Table. \ref{tab:Learned and True STLs} as a constraint, the policy learning curve of the TD3-Lagrangian RL algorithm is shown in Fig. \ref{fig:policylearningcurve}. Supplementary to Fig. \ref{fig:policylearningcurve}, we provide numeric values of key performance metrics: cumulative reward per episode and the cumulative cost per episode in Table. \ref{tab: numerical values for curve}. These metrics are derived from the data collected at the conclusion of the training phase averaged over three runs with three random seeds. 

\begin{figure*}[htbp]
\centering
\begin{subfigure}{0.33\textwidth}
  \includegraphics[width=\linewidth]{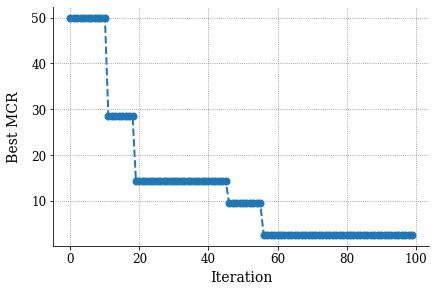}
  \caption{Point Circle}
  \label{fig:bospc}
\end{subfigure}%
\hfill
\begin{subfigure}{0.33\textwidth}
  \includegraphics[width=\linewidth]{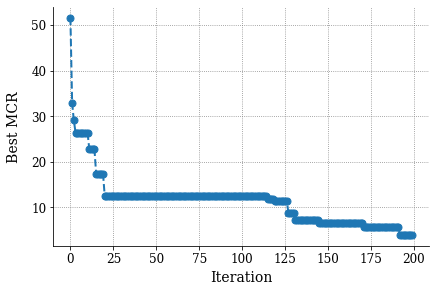}
  \caption{Point Goal}
  \label{fig:bospg}
\end{subfigure}%
\hfill
\begin{subfigure}{0.33\textwidth}
  \includegraphics[width=\linewidth]{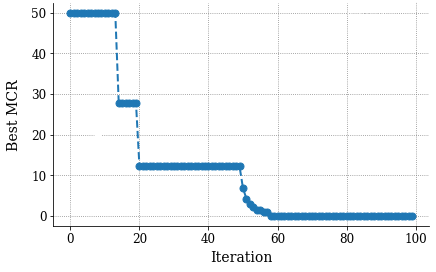}
  \caption{Half Cheetah Velocity}
  \label{fig:boshcv}
\end{subfigure}
\caption{BO learning curve for parameter learning of pSTL specifications provided in case studies. \ref{fig:bospc} depicts the learning curve for optimizing two parameters, \ref{fig:bospg} shows the learning curve for optimizing 16 parameters, and \ref{fig:boshcv} depicts the learning curve for optimizing two parameters. The minimization metric is given as the balanced misclassification rate (MCR) of the STL at sequentially generated candidate points. }
\label{fig:BO}
\end{figure*}

\begin{table*}[htpb]
\centering
\caption{Learned STL specifications alongside True environmental safety constraint for each case study}
\begin{tabularx}{\linewidth}{@{}c *5{>{\centering\arraybackslash}X}@{}}
\toprule
 & \textbf{Learned STL Specification ($\phi_{cost}$)} & \textbf{True STL Specification ($\phi_{true}$)} \\ \midrule \midrule

\makecell{\textbf{Safe Navigation} \\ \textbf{Circle}} &
$ G\left(\lnot\left((x_{a} < \text{-}0.93 ) \vee (x_{a} > 1.064)\right)\right) $
&
$ G\left(\lnot(\left(x_{a} < \text{-}1.20) \vee (x_{a} > 1.0 )\right)\right) $     
\\
\midrule
\makecell{\textbf{Safe Navigation} \\ \textbf{Goal}} &
$ \begin{aligned}
G \Biggl(\neg \biggl( 
&\biggl( \sqrt{{(x_{a}-\text{-}0.714)^2+(y_{a}-0.91)^2}} < 0.4 \biggr)\\
& \lor \biggl( \sqrt{{(x_{a}- 1.07)^2+(y_{a}-0.04)^2}} < 0.4 \biggr) \\
& \lor \biggl( \sqrt{{(x_{a}-\text{-}1.51)^2+(y_{a}-0.11)^2}} < 0.4 \biggr) \\
& \lor \biggl( \sqrt{{(x_{a}- \text{-}0.53)^2+(y_{a}- \text{-}0.31)^2}} < 0.4 \biggr) \\
& \lor \biggl( \sqrt{{(x_{a}- 0.43)^2+(y_{a} -0.87)^2}} < 0.4 \biggr) \\
& \lor \biggl( \sqrt{{(x_{a} - 0.16)^2+(y_{a} - \text{-}1.79)^2}} < 0.4 \biggr) \\
& \lor \biggl( \sqrt{{(x_{a}- \text{-}2.3)^2+(y_{a} -1.04)^2}} < 0.4 \biggr) \\ 
& \lor \biggl( \sqrt{{(x_{a}-0.83)^2+(y_{a}- \text{-}1.01)^2}} < 0.4 \biggr) \biggr) \Biggr) \end{aligned} \nonumber $
&
$\begin{aligned} G \Biggl(\neg \biggl(
& \biggl( \sqrt{{(x_{a}-\text{-}0.75)^2+(y_{a}-1.0)^2}} < 0.4 \biggr) \\
& \lor \biggl( \sqrt{{(x_{a}- 1.0)^2+(y_{a}-0.2)^2}} < 0.4 \biggr) \\
& \lor \biggl( \sqrt{{(x_{a}-\text{-}1.4)^2+(y_{a}-0.7)^2}} < 0.4 \biggr) \\
& \lor \biggl( \sqrt{{(x_{a}- \text{-}0.5)^2+(y_{a}- \text{-}0.3)^2}} < 0.4 \biggr) \\
& \lor \biggl( \sqrt{{(x_{a}-0.25)^2+(y_{a} -0.9)^2}} < 0.4 \biggr) \\
& \lor \biggl( \sqrt{{(x_{a} - 0.0)^2+(y_{a}- \text{-}1.5)^2}} < 0.4 \biggr) \\
& \lor \biggl( \sqrt{{(x_{a}- \text{-}1.9)^2+(y_{a} -1.0)^2}} < 0.4 \biggr) \\
& \lor \biggl( \sqrt{{(x_{a}-1.0)^2+(y_{a}- \text{-}1.0)^2}} < 0.4 \biggr) \biggr) \Biggr)\end{aligned} \nonumber $

\\
\midrule
\makecell{\textbf{Safe Velocity} \\ \textbf{Half Cheetah}} &
 $G \left(\neg \left(  u_{a} > 3.3521 \right)\right)) \nonumber $
&
$ G \left(\neg \left(  u_{a} > 3.2096 \right)\right)) \nonumber$
\\
\bottomrule
\end{tabularx}
\label{tab:Learned and True STLs}
\end{table*}

\begin{figure*}[!h]
\centering
\begin{subfigure}{\textwidth}
   \centering
   \includegraphics[width=0.8\linewidth]{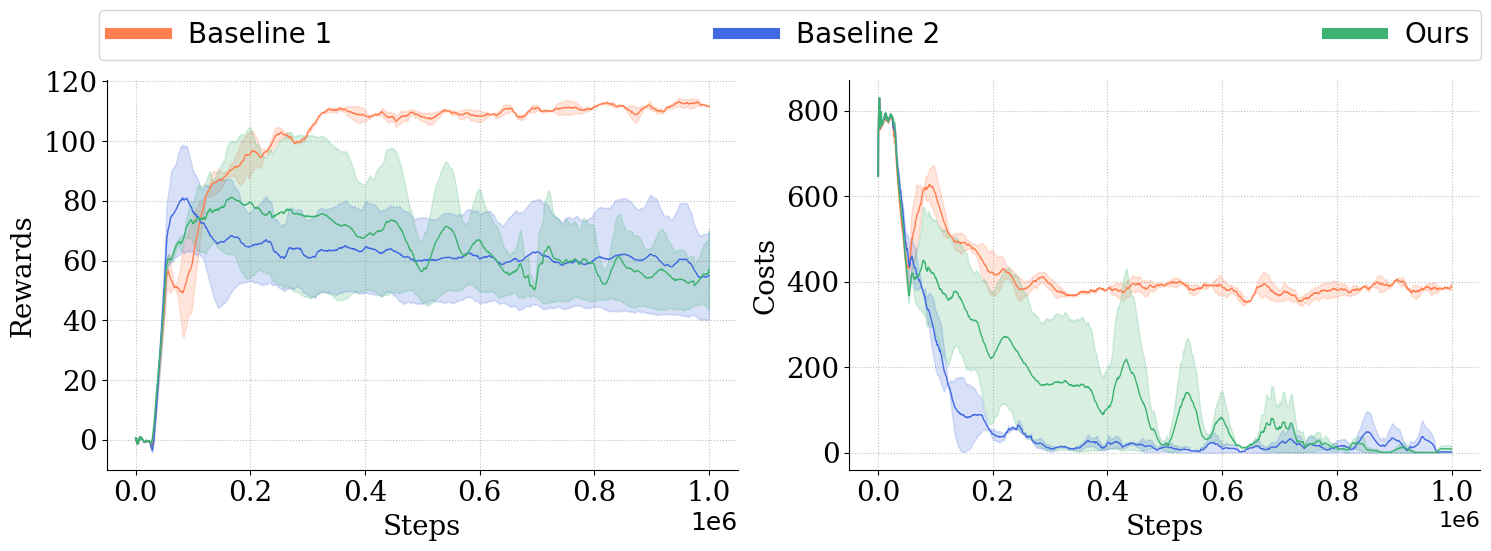}
   \caption{Safe Navigation - Circle}
   \label{fig:sub1}
\end{subfigure}

\vspace{0.1cm} 

\begin{subfigure}{\textwidth}
   \centering
   \includegraphics[width=0.8\linewidth]{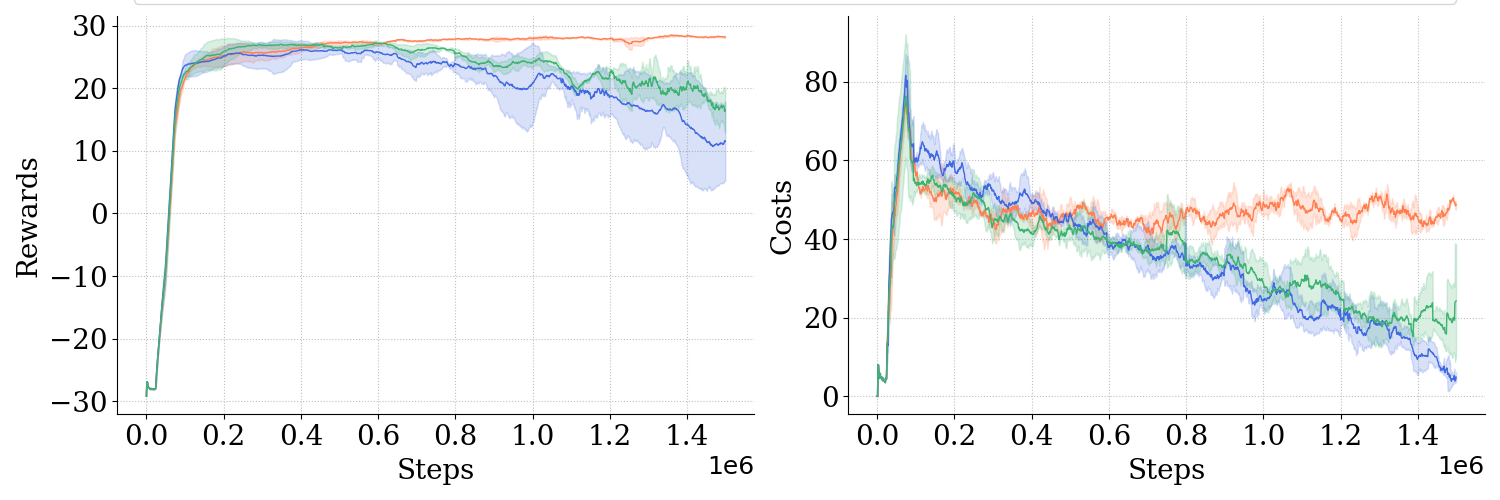}
   \caption{Safe Navigation - Goal}
   \label{fig:sub2}
\end{subfigure}

\vspace{0.1cm} 

\begin{subfigure}{\textwidth}
   \centering
   \includegraphics[width=0.8\linewidth]{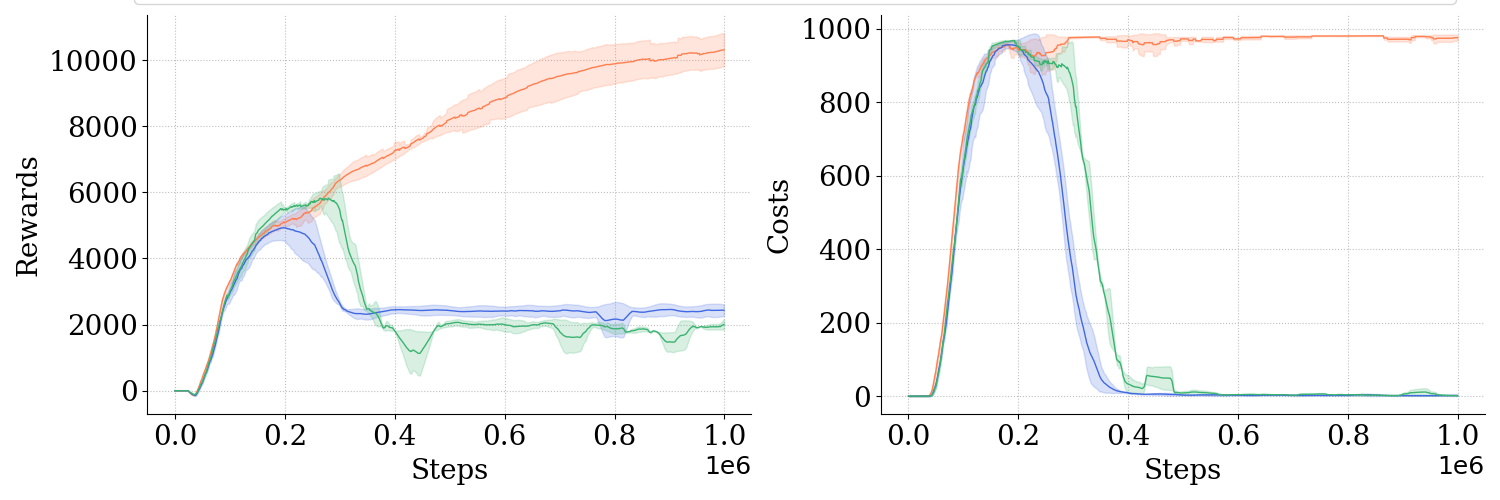}
   \caption{Safe Velocity - Half Cheetah}
   \label{fig:sub3}
\end{subfigure}

\caption{Policy learning curve for our algorithm and baselines 1 and 2 over $1e6$ total environment interactions for \ref{fig:sub1} and \ref{fig:sub3} and $1.5e6$ total environment interactions for \ref{fig:sub2}. The plots on the right display the cumulative rewards per episode, while those on the left display the cumulative costs (quantified as the total number of constraint violations) per episode throughout the training.}
\label{fig:policylearningcurve}
\end{figure*}

\begin{table}[htbp]
\caption{Metrics from the conclusion of training averaged over three random seeds per environment}
\label{tab: numerical values for curve}
\centering
\begin{tabularx}{\linewidth}{@{}c*{5}{X}c@{}} 
\toprule 
 & \multicolumn{2}{c}{\textbf{Baseline 1}} & \multicolumn{2}{c}{\textbf{Baseline 2}} & \multicolumn{2}{c}{\textbf{Ours}} \\
\cmidrule(lr){2-3} \cmidrule(lr){4-5} \cmidrule(lr){6-7}  
& $\mathcal{\overline{J}}_R$ & $\mathcal{\overline{J}}_c$ & $\mathcal{\overline{J}}_R$ & $\mathcal{\overline{J}}_c$ & $\mathcal{\overline{J}}_R$ & $\mathcal{\overline{J}}_c$ \\
\midrule \midrule
\makecell{\textbf{Safe Navigation} \\ \textbf{Circle}} & 111.3 & 390.3 & 54.90 & 1.41 & 57.02 & 8.39 \\
\midrule
\makecell{\textbf{Safe Navigation} \\ \textbf{Goal}}  & 28.2 & 48.8 & 11.5 & 4.9 & 16.5 & 24.3 \\ 
\midrule
\makecell{\textbf{Safe Velocity} \\ \textbf{Half Cheetah}} & 10371.1 & 957.6 & 2676.1 & 1.67 & 2114.7 & 0.62 \\ 
\bottomrule
\end{tabularx}
\end{table}

Subsequently, we display a graphical illustration depicting the percentage of safe traces within a set of rollout traces generated by executing the trained policy across each case study. This analysis includes a comparison between baseline 1, baseline 2, and our implementation on the various case studies and is depicted in Fig. \ref{fig:bargraph}. The primary objective of this evaluation is to provide a quantifiable measure of safety for policies generated through each approach, effectively gauging the potential rate of unsafe incidents that might occur if any of these policies were to be deployed in the respective case studies. Such a visual and statistical comparison is instrumental in assessing the relative safety efficacy of each approach.
\begin{figure}[htbp]
\centering
    \includegraphics[width=\linewidth]{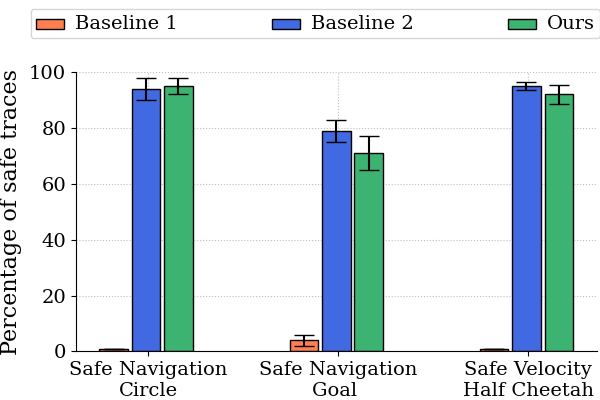}
    \caption{Percentage of safe traces in final policy rollouts across different case-studies implementing proposed and baseline algorithms.}
    \label{fig:bargraph}
\end{figure}

In our final analysis, we conduct a comprehensive evaluation focusing on the performance of the upper-level optimization, specifically the learning of the STL constraint parameters. This evaluation entails calculating the misclassification rate of the STL post-training against a labeled data set to assess the accuracy of our learned STL. In order to establish a benchmark for this metric, we compare it with that of the MCR of STL constraint used in baseline 2, which is the true environmental constraint, against the same dataset. A close alignment in these rates would indicate a high degree of accuracy of our learned parameters relative to the \emph{true} parameters.

\begin{table}[htbp]
\caption{MCR comparison between the learned STL (ours) and the true STL (baseline 2)}
\label{tab2}
\centering
\begin{tabular}{@{}c*{6}{c}}
\toprule 
&\multicolumn{2}{c}{\textbf{MCR}}\\
\cmidrule(lr){2-3}
&{Baseline 2} & Ours \\
\midrule \midrule
\makecell{\textbf{Safe Navigation} \\ \textbf{Circle}} & 0.0 & 0.0251  \\
\midrule
\makecell{\textbf{Safe Navigation} \\ \textbf{Goal}}  & 0.0 & 0.0534  \\ 
\midrule
\makecell{\textbf{Safe Velocity} \\ \textbf{Half Cheetah}} & 0.0 & 0.0  \\ 
\bottomrule
\end{tabular}
\label{tab: MCR}
\end{table}

In the analysis presented in Fig. \ref{fig:policylearningcurve}, a trade-off between rewards and costs is observed across all case studies. This observation substantiates that the tasks in all of the case studies are not \enquote{trivially-safe}, i.e., maximizing rewards in these settings consistently leads to constraint violations to some degree. Notably, baseline 1 achieves the highest reward in all case studies, yet it concurrently incurs the highest cost at the end of training. This pattern suggests that the agent continues to engage in unsafe actions, prioritizing only reward maximization. In contrast, our algorithm exhibits a reduction in rewards compared to baseline 1; however, it succeeds in reducing costs substantially across all case studies, and even achieves the threshold of zero violations per episode by the end of training in two out of the three case studies. This improvement upon baseline 1 is a direct result of applying our algorithm in scenarios with initially unknown safety constraints by allowing the learning and adhering to safety constraints, even in the absence of prior knowledge of the constraints. Baseline 2 represents an ideal training scenario, assuming complete availability of STL safety constraint information. The performance of our algorithm closely mirrors that of baseline 2, a result that indicates close similarity between the learned STL in our approach and the true STL. 

The data provided in Table. \ref{tab: numerical values for curve} offers a quantitative counterpart to the results depicted in Fig. \ref{fig:policylearningcurve}. This tabular representation offers a more detailed numerical articulation of the result metrics at the end of training, complementing the plots displayed in the figure. Consequently, the interpretation of the results in the table aligns closely with that of Fig. \ref{fig:policylearningcurve}, i.e. while baseline 1 achieves the highest cumulative rewards per episode at the conclusion of training, it also incurs the highest number of constraint violation per episode. In contrast, the adoption of our proposed method demonstrates a significant improvement in cost-efficiency by efficiently directing the agent to act in accordance to the learned environmental constraints, which closely mirror the actual environmental constraints. 

In Fig. \ref{fig:bargraph}, it is evident that the policy optimized under baseline 1 fails to produce safe trajectories in case studies 2 and 3, with only a few safe trajectories in case study 2. In contrast, the policy optimized through our framework yields a number of safe trajectories comparable to baseline 2, which had complete knowledge of the safety constraints from the start. This demonstrates the effectiveness of our approach in learning the safety constraints and ensuring safety during policy deployment, even with less initial information on safety constraints.

In Table. \ref{tab: MCR}, we exhibit results that underscore the quality of the learned STL using our approach. We assessed the STL's quality by its ability to accurately classify labeled data, and then benchmarked these results against the performance of the true STL used in baseline 2. While the true STL safety specification, by definition, should classify all traces with a misclassification rate (MCR) of zero, it is noteworthy that the MCR of the STL derived through our algorithm closely parallels this standard. In scenarios such as case studies 1 and 3, characterized by a limited number of learning parameters for the pSTL, the MCR is close to zero, mirroring the performance of the true STL, whereas in more complex settings, such as that of case study 2 with 16 learning parameters, the MCR, while higher, still remains within reasonable bounds considering the large number of learning parameters. This not only highlights the precision of our STL learning process but also indicates that the parameters we derived are remarkably close to the real environmental constraints. Overall, our results demonstrate the precision of our algorithm  in adapting to and respecting the environmental safety constraints, thereby offering a balanced approach in terms of performance and cost during training and implementation.

\noindent{\textbf{Limitations.}} Despite the successful results, there are limitations to our approach which must be acknowledged. Firstly, our approach relies on pre-existing datasets of safe and unsafe trajectories, however small, as well as an STL safety specification template. The  availability of these elements is required for the initialization of our process and the overall performance. The second limitation is the requirement for human expert manual labeling of trajectories. While human expertise is invaluable for providing a better understanding of safety, this requirement imposes considerable demands on human resources. Finally, despite the complex integration of our framework, it does not guarantee the derivation of a safe policy.


\section{Conclusion}\label{sec:Conclusion}
This research tackles the challenge of ensuring safety in RL, particularly when predefined safety constraints are unavailable. Traditional methods in safe RL often rely heavily on static, predefined safety constraints, thus limiting their applicability. To address this limitation, we proposed an approach that concurrently learns an optimal control policy and identifies the STL safety constraint parameters of a given environment. Our approach implements a bilevel optimization framework, where the upper level is dedicated to optimizing parameters of pSTL safety constraint, and the lower level aims to find an optimal safe policy, constrained by the learned STL safety specification. Our process also leverages input from human experts who assign safety labels to the RL policy rollout traces to be used to refine safety specification parameters. Various case studies demonstrate the efficacy of our approach, showing that our algorithm substantially reduces constraint violations compared to traditional unconstrained reward maximization methods, while maintaining similar levels of performance. Additionally, it closely mirrors the results of scenarios with complete initial knowledge of true environmental constraints, thereby underscoring the close alignment of our learned STL parameters with actual safety parameters. 

\bibliographystyle{ieeetr}
\bibliography{main}

\end{document}